\documentclass[aps,prl,reprint,footinbib,floatfix]{revtex4-1}
\usepackage{amsmath}
\usepackage{amssymb}
\usepackage{graphicx}

\usepackage[usenames]{color}
\usepackage[normalem]{ulem}

\def\be{\begin{equation}}
\def\ee{\end{equation}}
\def\ber{\begin{eqnarray}}
\def\eer{\end{eqnarray}}
\def\bern{\begin{eqnarray*}}
\def\eern{\end{eqnarray*}}

\def\rv{\mathbf{r}}

\def\Gv{\mathbf{G}}
\def\kv{\mathbf{k}}

\def\qv{\mathbf{q}}

\def\bv{\mathbf{b}}

\def\0v{\mathbf{0}}
\def\1v{\mathbf{1}}
\def\2v{\mathbf{2}}
\def\3v{\mathbf{3}}

\def\rp{\rv_{\scriptscriptstyle{\|}}}

\begin{document}

\title{Scattering resonances in graphene}

\author {V. U. Nazarov}
\affiliation{Research Center for Applied Sciences, Academia Sinica, Taipei 11529, Taiwan}
\email{nazarov@gate.sinica.edu.tw}

\author{E. E. Krasovskii}
\affiliation{Departamento de F\'{i}sica de Materiales, Facultad de Ciencias Qu\'{i}imicas, Universidad del Pais Vasco/Euskal Herriko Unibertsitatea, Apdo. 1072, San Sebasti\'{a}n/Donostia, 20080 Basque Country, Spain}
\affiliation{Donostia International Physics Center (DIPC), Paseo Manuel de Lardizabal 4, San Sebasti\'{a}n/Donostia, 20018 Basque Country, Spain}
\affiliation{IKERBASQUE, Basque Foundation for Science, 48011 Bilbao, Spain}

\author{V. M. Silkin}
\affiliation{Departamento de F\'{i}sica de Materiales, Facultad de Ciencias Qu\'{i}imicas, Universidad del Pais Vasco/Euskal Herriko Unibertsitatea, Apdo. 1072, San Sebasti\'{a}n/Donostia, 20080 Basque Country, Spain}
\affiliation{Donostia International Physics Center (DIPC), Paseo Manuel de Lardizabal 4, San Sebasti\'{a}n/Donostia, 20018 Basque Country, Spain}
\affiliation{IKERBASQUE, Basque Foundation for Science, 48011 Bilbao, Spain}

\begin{abstract}
We address the two-dimensional band-structure of graphene  above the vacuum level 
in the context of discrete states immersed in the
three-dimensional continuum. Scattering resonances are discovered
that originate from the coupling of the in-plane and 
perpendicular motions, as elucidated by the analysis of an exactly solvable 
model. Some of the resonances turn into true bound 
states at high-symmetry $\kv$ vectors. {\it Ab initio} scattering 
theory verifies the existence of the resonances in realistic graphene 
and shows that they lead to a total reflection of the incident electron 
below and total transmission above the resonance energy.
\end{abstract}


\maketitle
Electronic structure of single layer crystals has attracted much 
attention due to the discovery of graphene~\cite{Neto-09} and other 
atomically thin systems (boron nitride~\cite{BN10}, silicene~\cite{silicene12}). 
Graphene is the most popular material because it combines the unique 
electronic properties with technological robustness, which makes it 
especially promising for nanoelectronics~\cite{Ponomarenko08}. Its 
most exciting feature -- the linear dispersion of the highest occupied 
$\pi$ and lowest empty $\pi^*$ bands -- is known since 1947, when 
Wallace~\cite{Wallace-47}  obtained it analytically in a 
tight-binding model. The bound electronic states of the free-standing 
graphene have been recently addressed in a number of {\it ab initio}
studies~\cite{Latil-06,Trevisanutto-08,Malard-09,Silkin-09,Suzuki-10,Kogan-12},
so its low energy band structure is presently well understood.

\begin{figure}[h] 
\includegraphics[width=1.0 \columnwidth,clip=true,trim=30 0 0 0]{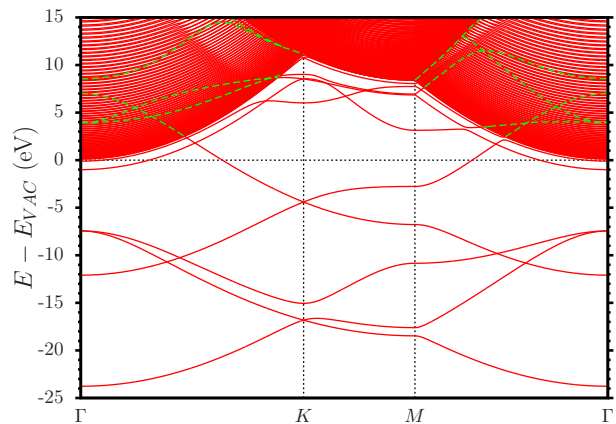} 
\caption{\label{fig_graphene} (color online) The band-structure of
  graphene obtained in the repeated super-cell geometry. The
  all-electron full-potential linearized augmented-plane wave code
  {\it Elk} \cite{Elk} was used for this calculation. The separation
  between the periodically stacked layers is $d=400$ bohr. The energy axis zero is at the vacuum level at $\Gamma$ point.}
\end{figure}

At energies above the vacuum level we enter the continuous spectrum
due to the infinite motion perpendicular to the layer, as shown in
Fig.~\ref{fig_graphene} for the band-structure of graphene.  Some of
the lines entering the continuum from below are seen to retain their
individuality inside the continuous spectrum.  Their origin is clear:
They correspond to the states of an in-plane motion but with the
energy above the continuum edge.  However, at a deeper
level, a fundamental issue arises: An electron moving with a
sufficiently high energy within the layer and parallel to it has,
generally speaking, a non-zero probability to escape into vacuum,
which would impart this state a finite life-time, i.e., turn it into a
resonance. The presence of the resonances in the band-structure of graphene
above the vacuum level is the main message of this Letter. 
We argue that
those resonances  are of special kind: They originate from
the  coupling of two motions, of
which one is across the layer under the action of the layer's confining
potential well and another is in the layer's plane in
a periodic 2D lattice potential, {\em each of those potentials
separately supporting no resonances}. We will also show that
some of the discrete levels retain zero linewidth, by this being true
bound states immersed into continuum.

The discrete levels within the continuous spectrum deserve close 
attention because they strongly affect optical 
absorption in the UV range, as well as electron photoexcitation and propagation 
toward the detector in a photoemission experiment. Recent experimental progress 
in angle-resolved photoemission (ARPES)
on epitaxial~\cite{Shikin00,Bostwick06,Dedkov08,Sutter09,C-Ir09,Liu10} 
and suspended~\cite{Niesner12,Knox11}  
graphene, as well as in low energy electron diffraction
(LEED)~\cite{Locatelli10,Sutter08} calls for a detailed understanding of 
its electronic structure at higher energies.

We start by considering a trivial case: 
Let us have a quantum well $V(z)$  
in $z$ direction with the flat potential in $xy$ plane. Then, if the
well supports a bound state, and since the two perpendicular motions
are independent in this case, the 3D wave-function is the product of a
bound state in $z$ direction and a plane wave in the $xy$ plane.  As a
result, there exist states which are bound to the well while having an
arbitrarily high energy above the vacuum level due to the motion in
$xy$ plane.
If, however, we apply a potential that 
is periodic in $xy$ plane, the variables in the  Schr\"{o}dinger equation 
do not separate any more, i.e., the two perpendicular motions become coupled.
To get a better insight on how this affects the high-lying energy bands of the 
in-plane motion, we first introduce a model which is exactly solvable and at the same time retains
all the basic physics involved:

{\it $\delta$-function quantum well with laterally periodic potential --}
We are looking for a solution of the  Schr\"{o}dinger equation 
\begin{equation}
\left[-\frac{1}{2} \Delta + V(z,\rp) \right] \psi(z,\rp) = E \psi(z,\rp)
\label{schr}
\end{equation}
with the model potential being a product of a periodic function in the $xy$ plane and the $\delta$-function
quantum well in $z$ direction 
\begin{equation}
\label{V}
V(z,\rp)= \sum\limits_{\Gv} V_\Gv e^{i \Gv\cdot\rp} \, \delta(z),
\end{equation}
where $\Gv$ are the 2D reciprocal lattice vectors.
We set $V_{\0v}<0$ to ensure the existence of a state bound to the $z=0$ plane.
The solutions of Eq.~(\ref{schr}) with the potential (\ref{V}) can be written explicitly as Bloch waves
with respect to the motion in the $xy$ plane
\begin{equation}
\psi(z,\rp) = \sum\limits_{\Gv} a_{\Gv} e^{i \sqrt{2 E-(\Gv+\kv)^2} \,|z|} e^{i (\Gv+\kv)\cdot\rp},
\label{sol}
\end{equation}
where 
$\kv$ is the in-plane wave-vector within the first Brillouin zone,
and $a_\Gv$ are still unknown coefficients.
Importantly, in Eq.~(\ref{sol}) we have retained the exponent with one sign only,
which selects out
bound and resonant states \cite{Landau-81}, if the latter exist,  
while omitting the scattering states propagating in the $z$ direction
\footnote{All the solutions (\ref{sol}) are even in $z$ ($\sigma$ bands),
which is the consequence of the $\delta$-potential supporting one bound state at most.}.
The jump in the wave-function's $z$-derivative $\psi'(z,\rp)$
is obtained by the integration of Eq.~(\ref{schr}) in $z$ over 
the infinitesimal interval $[0_-,0_+]$
\begin{equation}
\psi'( 0_+,\rp) \! - \! \psi'( 0_-,\rp) \! = \! 2 \sum\limits_{\Gv} V_\Gv e^{i \Gv\cdot\rp} \psi( 0,\rp).
\label{b}
\end{equation}
Together, Eqs. (\ref{sol}) and (\ref{b}) lead to the system of equations for 
the coefficients $a_{\Gv}$
\begin{equation}
\sum\limits_{\Gv'} V(\Gv  -  \Gv') a_{\Gv'}  =  i \sqrt{2 E- (\Gv  +  \kv)^2  } \, a_{\Gv}.
\label{eq}
\end{equation}

The crucial point is the choice of the sign of the square roots in Eqs.~(\ref{sol})
and (\ref{eq}). 
Denoting the generic square root by $s$, the rule is:
${\rm Re}\ s >0$ if
${\rm Re}\, s^2 >0$  and ${\rm Im}\ s >0$ otherwise,
which choice ensures the correct asymptotic behavior 
of the necessarily normalizable and the necessarily non-normalizable wave-functions,
of the bound states and resonances, respectively \cite{Landau-81}. 

The values of $E$ which allow for non-zero 
solutions of the homogeneous system of linear equations (\ref{eq}) 
determine the band-structure of our model system.  However, in
contrast to the original set-up of Eq.~(\ref{schr}), Eqs.~(\ref{eq})
constitute a {\it nonlinear eigenvalue problem} \cite{Golub-00} 
for the energies $E$.  We emphasize that 
this fundamental difference comes from the fact that the separation of 
the bound and resonant states from those of continuum has been already 
achieved in Eq.~(\ref{eq}).

\begin{figure}[h] 
\includegraphics[height=0.52 \columnwidth,clip=true,trim=40 0 46 0]{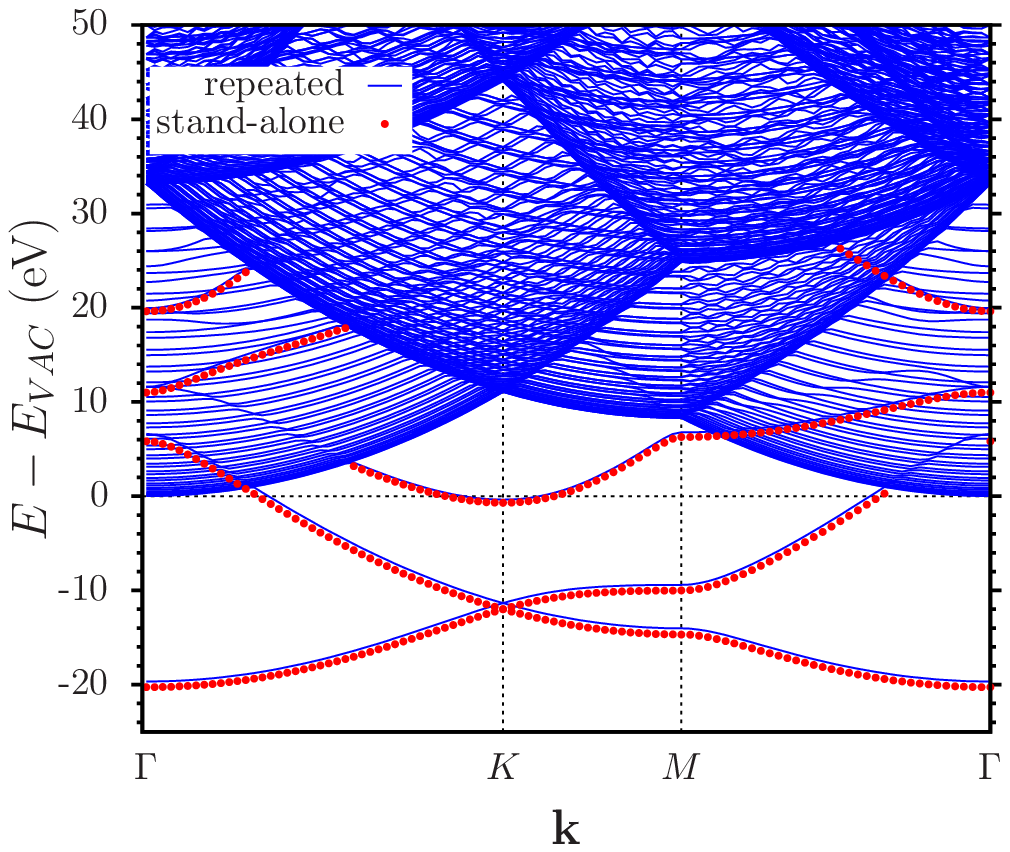} 
\includegraphics[height=0.52 \columnwidth,clip=true,trim=95 0 86 0]{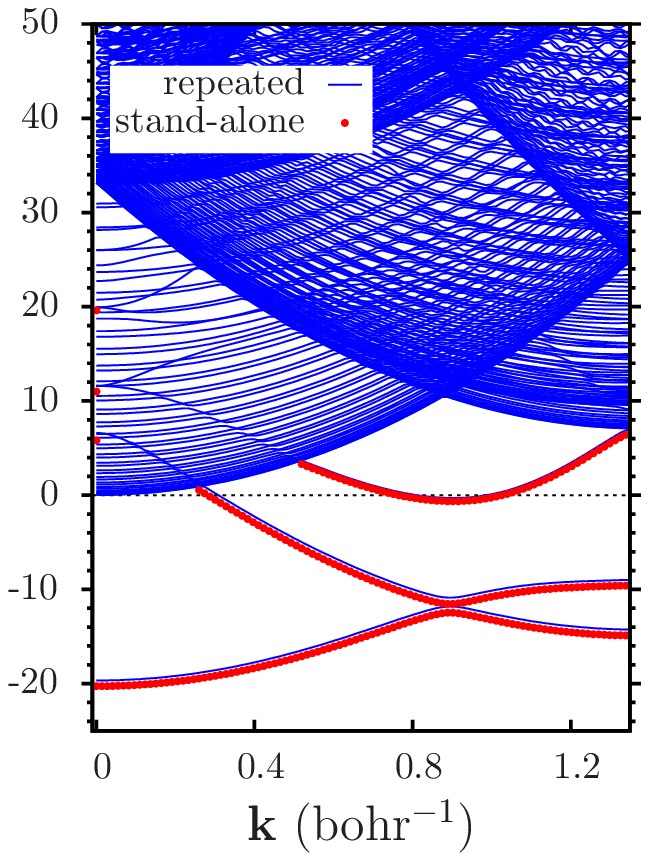}
\caption{\label{fig_model} (color online)
Left: Band-structure of the model system obtained with the repeated super-cell geometry calculation (blue lines) and with solving the eigenvalue problem (\ref{eq}) for a stand-alone plane (red points). The calculation has been conducted along $\Gamma-K-M-\Gamma$ line.
Right: The same for the asymmetric  direction of $\kv$ along $7\bv_1+13\bv_2$, where $\bv_1$ and $\bv_2$ are the primitive reciprocal vectors.
The energy axis zero is at the vacuum level at $\Gamma$ point.
}
\end{figure}

In Fig. \ref{fig_model}, results of the numerical solution of the
non-linear eigenvalue problem (\ref{eq}),  which give the
band-structure of the stand-alone layer,  are presented
together with the results of calculations carried out for the same
system in the repeated super-cell geometry.
The symmetry of the 2D periodic potential is chosen that 
of the honeycomb lattice, graphene's lattice constant is used, 
and the values of
the Fourier coefficients of the potential in atomic units are:
$V_\0v=-0.7 |\bv|$ and $V_\Gv=0.1 |\bv|$ for the first $\Gv$ star, and
$V_\Gv=0$ otherwise, $\bv$ being the primitive
vector of the reciprocal lattice.  In the left panel, the wave-vector varies along
the $\Gamma-K-M-\Gamma$ lines while in the right panel 
an asymmetric direction of $\kv$ is chosen. For the bands 
below the vacuum edge, both the repeated-geometry and the stand-alone 
calculations yield the identical results regardless of the symmetry of 
the wave-vector. In contrast, above the vacuum edge, whether or not a 
particular state localized near $z=0$ survives as a true bound state is 
determined by the symmetry of its $\kv$ point. For the 
asymmetric case of $\kv$,
there are no such states. As can be seen from Fig.~\ref{fig_model}, along the 
high-symmetry directions some of the bound state bands 
do survive. Moreover, an isolated high-lying bound state exists at the $\Gamma$ point
at the energy of $\approx$83 eV (not shown in Fig.~\ref{fig_model}). 
The 2D crystal, thus, presents a simple and instructive 
example of bound states in the continuum, very different from the known cases 
of atoms~\cite{FW85} or quantum dots~\cite{CL08}.

While the results in Fig.~\ref{fig_model} establish the existence
of bound states  above the vacuum edge, they do
not answer the question of what happens with those 
that do not survive, i.e., do the latter turn into resonances by 
acquiring a finite lifetime or they disappear 
at all. This is due to our numerical search for the eigenvalues of the 
nonlinear eigenproblem (\ref{eq}) having been restricted to the real axis 
of $E$, since no decisive  
numerical procedure exists to either find all complex-valued roots of
this problem or to prove their absence. To shed  light on that issue, we solve
the  eigenvalue problem (\ref{eq}) analytically for a reduced size
of the $V(\Gv-\Gv')$ matrix to make the problem computationally feasible.
Using the  {\it Mathematica} symbolic algebra software, we have 
analytically evaluated the determinant $\Delta(E)$ of the 
system (\ref{eq}), then consecutively eliminated the square 
roots in the equation $\Delta(E)=0$, which 
made it possible to reduce it 
to a polynomial equation. 
All roots of the polynomial (including the complex ones) 
were then found 
with no loss of any of them guaranteed. Since spurious zeros were introduced when
reducing the equation to the polynomial, the roots were finally sorted to retain
only those that satisfied 
the original equation $\Delta(E)=0$.  This has been done for $\Gamma$, $K$, and $M$ 
points with the matrix sizes of 19, 13, and 7, respectively. 
While all true bound states, both 
below and above the vacuum edge, were found to reproduce 
those previously obtained numerically, in addition, complex eigenvalues were
found. Results of this calculation are collected in Table \ref{tab}.

\begin{table}
\caption{\label{tab}
Eigenenergies (in eV) for the model system
obtained with the reduced size of $V(\Gv-\Gv')$ matrix permitting the fully analytical solution
of the non-linear eigenvalue problem (\ref{eq}).}
\begin{tabular}{|l|l|l|}
\hline\hline
$\Gamma$ & $K$ & $M$ \\   
\hline
-20.3 & -11.9 & -13.9 \\
5.8 & -0.6 & -9.5 \\
11.0 & 26.0 - 0.2 i & 7.7 \\
19.6 & 26.8 - 0.6 i & 9.8 - 0.6 i \\
24.3 - 1.4 i & 28.5 - 0.9 i & 39.4 - 0.7 i \\ 
80.1 - 0.9 i & 55.9 - 0.2 i & 42.1 - 0.3 i \\ 
80.9 - 0.6 i & 61.6 - 0.6 i & 59.6 - 0.8 i \\
82.4 - 0.2 i & 61.6 - 0.7 i & \\
83.2 & 64.6 - 0.8 i & \\ 
116.5 - 0.2 i & 127.7 - 0.6 i & \\
116.8 - 0.6 i & 127.7 - 0.5 i & \\
117.3 - 1.3 i & 160.8 - 0.5 i & \\ 
\hline\hline
\end{tabular}
\end{table}

\begin{figure}[t]     
\includegraphics[width=0.95 \columnwidth,clip=true,trim=40 0 0 0]{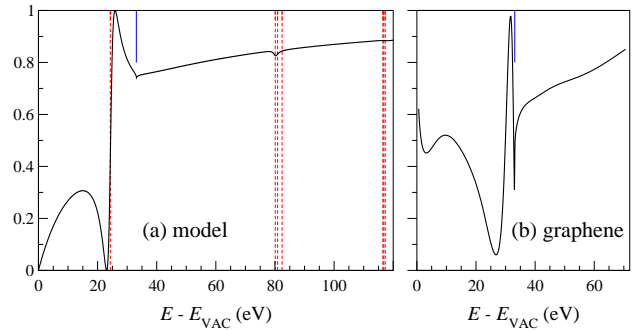}
\caption{\label{scat} (Color online)
Energy dependence of the electron transmission coefficient $T(E)$ through a 
free standing layer for the model system (a) and for graphene (b). The incidence 
is normal to the layer. Vertical dashed lines in graph (a) 
indicate the positions of resonances at $\Gamma$ point from Table~\ref{tab}.
In both graphs, vertical bars at kinetic energy of 33.1~eV indicate the onset 
of non-specular reflected beams.
}
\end{figure}

Eigenenergies in the lower complex half-plane (resonances)
physically manifest themselves as features in elastic scattering spectra
at the real energies in the vicinity of the complex eigenvalues \cite{Landau-81}.
In Fig. \ref{scat}(a) we plot the coefficient
of transmission of an electron incident normally onto our model system.
The features in the transmission spectrum clearly agree with the
resonances' positions listed in the first column of Table \ref{tab} ($\Gamma$ point).
We note that it is resonances, not the bound states, that
underlie the singularities in the elastic scattering spectra: Bound
states are orthogonal to the scattering states leading to the
independence of the two corresponding motions.

\begin{figure*}[th*]    
\includegraphics[height=0.3\textwidth]{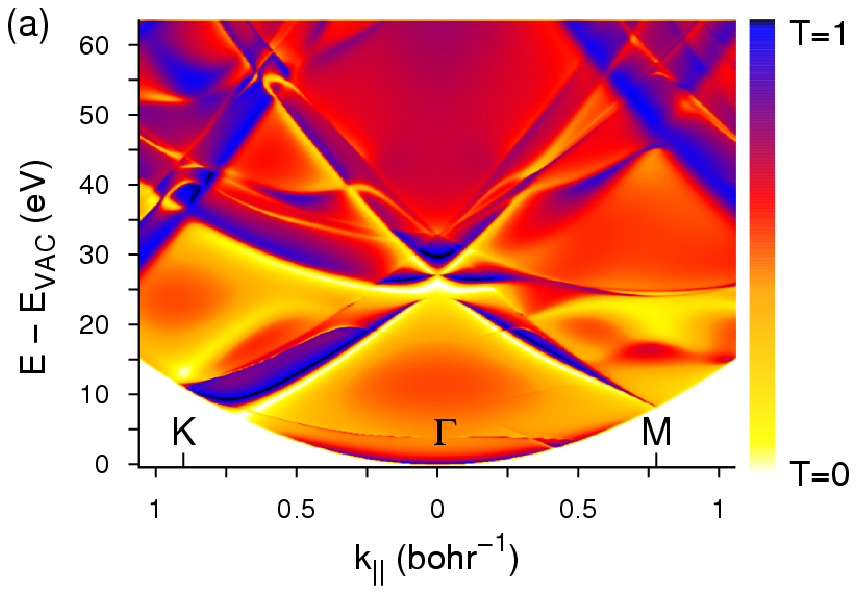}\hspace{5mm}
\includegraphics[height=0.3\textwidth]{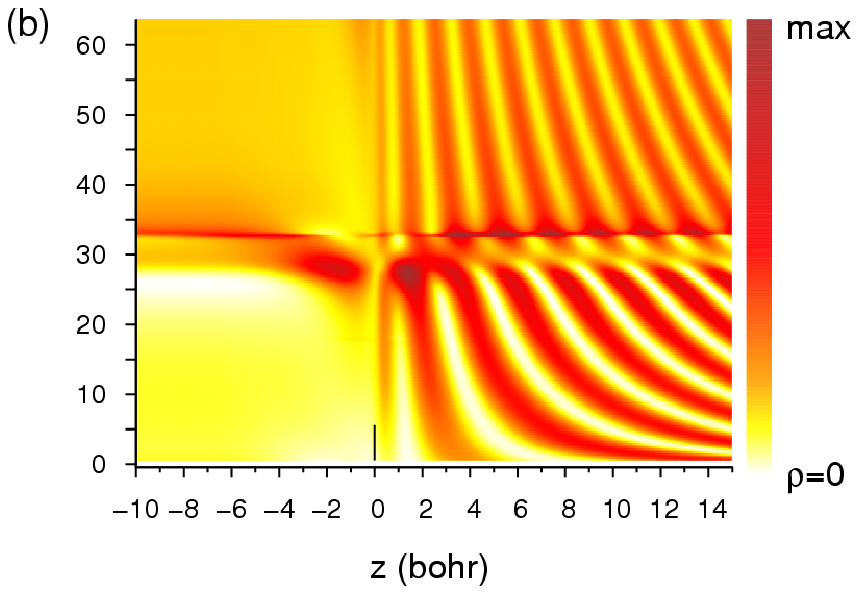}
\caption{\label{tcs} (Color online) (a) Energy-momentum distribution of the 
transmission probability $T(\kv,E)$ through the graphene 
monolayer. (b) Energy-dependent density distribution $\rho(z,E)$ in the 
normal incidence LEED state. Graphene plane is at $z=0$. 
The wave is incident from the right; it is normalized as $\exp(i \qv\cdot\rv)$. }
\end{figure*}                    
%

Having established the origin of high-energy resonances in  the infinitely thin system
let us now return to a realistic graphene. Experimentally, the scattering resonances 
can be observed in low energy electron diffraction. Figure~\ref{scat}(b) 
shows {\it ab~initio} normal incidence electron transmission spectrum calculated 
with the augmented-plane-waves (APW) based variational embedding method~\cite{K04}
\footnote{The method employs eigenfunctions of a repeated-slab band structure as basis
functions to represent the LEED state. For the present purpose they were 
obtained with the full-potential linear APW of Ref.~\cite{KSS99}.}.
The {\it ab initio} spectrum of graphene is similar to that of the 
model system: Just below the lowermost resonance we find
a point of total reflection followed by total transmission 
just above the resonance. Total reflection from a free-standing monolayer is a 
rather counterintuitive finding: unlike the well known case of LEED from crystal 
surfaces, it is not caused by a gap in the energy spectrum of the semi-infinite 
substrate~\cite{Slater_LEED37}. Indeed, the electron can freely propagate 
in the vacuum half-space behind the graphene layer, and the reflection is 
solely due to the in-plane scattering.

In both $T(E)$ spectra, one can also see a sharp structure due to the 
emergence of the secondary beams. Both in the model and in the actual graphene 
it appears as a transmission minimum at the same energy $E_{\rm KIN}=33.1$~eV,
see Fig.~\ref{scat}. Such structures are well known in  classical 
LEED~\cite{McRae79,Jones88}, and, contrary to the ones found here, they have a 
purely ``structural'' origin and do not depend on details of the electronic 
structure.

By changing the incidence angle one can observe the {\em scattering band structure}
as the dispersion of transmission probability with $\kv$. The 
{\it ab initio} calculation of $T(\kv,E)$ in the directions
$\Gamma K$ and $\Gamma M$ is presented in Fig.~\ref{tcs}(a).
The resonance at $\Gamma$ is 
seen to split at the off-normal incidence into three 
branches with a pronounced anisotropic dispersion, which highlights 
the non-free-electron character of the graphene 
states at high energies. A high intensity of Umklapp bands is seen as 
well. To visualize the scattering resonance in  
real space we present in Fig.~\ref{tcs}(b) the energy dependence of the 
electron density distribution in the LEED state as it comes out of our 
{\it ab initio} calculation. The white stripe in the left half-space at 
$E_{\rm KIN}=25.5$~eV corresponds to the total reflection, and the vanishing 
beating in the right half-space at 31.5~eV to the total transmission. The 
resonance is seen as the pronounced local density enhancement at the graphene 
layer at 27.5~eV. In perfect accord 
with our model, it is located between the minimum and the maximum of $T(E)$.

The discovered resonances are, thus, typical of atomic monolayers, and
at the surfaces of 3D crystals they may be blurred by the interlayer scattering.
For example, in graphite, the resonance falls in a wide gap in the 
$\kv=0$ projected spectrum, see Fig.~6 in Ref.~\cite{BKTS05}. 

These findings suggest important implications on LEED and ARPES from graphene. 
The two techniques are related by the one-step theory \cite{Mahan70,FE74}, 
according to which the 
photocurrent is proportional to the probability of the optical transition to 
the time-reversed LEED state. For the supported graphene, for a sufficiently 
weak interaction with the substrate, one can, apparently, reduce or enhance 
the signal from the substrate by tuning the photon energy to the reflection 
or transmission point. The resonances are rather prominent also at off-normal 
incidence [Fig.~\ref{tcs}(a)]. As they are associated with a strong in-plane 
scattering it is especially important to be aware of them in studying the 
corrugated suspended graphene with LEED or ARPES because the resonance area 
is most strongly affected by the lattice deformation. 

To summarize, we have shown that atomically thin monolayers support
resonances of special nature: they originate from a strong coupling
of the in-layer scattering to the motion perpendicular to the 
layer, while each of the two motions, separately, does not support a 
resonance. For the exactly solvable model of an infinitely-thin
crystal we have found the complex eigenvalues of the resonances and 
demonstrated that they lead to strong sharp structures 
in the electron diffraction spectra. Another interesting result is 
that apart from the resonances there exist true bound states immersed
in the continuum spectrum, which survive up to high energies above the 
vacuum level. The purely real eigenvalues are, however, restricted to 
high-symmetry directions of the 2D Brillouin zone, and they turn into 
resonances at general $\kv$ points.

These general results have found full verification in our
{\it ab initio} calculation of electron diffraction from a realistic 
free-standing graphene monolayer. The resonance causes a total 
reflection of the normally incident electron with an energy just
below the resonance -- a unique phenomenon, as it is caused by purely
in-plain scattering. Such resonances are a general property of
atomic monolayers with vast implications for various fields 
of electron spectroscopy. 

Authors thank Eugene Kogan for valuable discussions.
V.U.N. acknowledges partial support from National Science Council, Taiwan,
Grant No. 100-2112-M-001-025-MY3.  E.E.K. and V.M.S. acknowledge partial
support from the Spanish Ministerio de Ciencia e Innovaci\'on (Grant
No. FIS2010-19609-C02-02).

\end{document}